\documentclass[11pt]{article}
\usepackage{aas_macros,amsmath,amssymb,color,graphics,epsfig}
\usepackage[numbers,square,comma,sort&compress,merge]{natbib}

\usepackage{amsfonts}
\usepackage{float}

\newcommand{\be}{\begin{equation}}
\newcommand{\ee}{\end{equation}}
\newcommand{\bea}{\setlength\arraycolsep{2pt} \begin{eqnarray}}
\newcommand{\eea}{\end{eqnarray}}

\newcommand{\mm}{\mathrm}

\def\ft#1#2{{\textstyle{\frac{\scriptstyle #1}{\scriptstyle #2} } }}
\def\fft#1#2{{\frac{#1}{#2}}}

\def\0{{\sst{(0)}}}
\def\1{{\sst{(1)}}}
\def\2{{\sst{(2)}}}
\def\3{{\sst{(3)}}}
\def\4{{\sst{(4)}}}
\def\5{{\sst{(5)}}}
\def\6{{\sst{(6)}}}
\def\7{{\sst{(7)}}}
\def\8{{\sst{(8)}}}
\def\sst#1{{\scriptscriptstyle #1}}

\begin{document}

\begin{flushright}
\end{flushright}

\vspace{25pt}
\begin{center}
{\large {\bf  Emergent symmetry and thermodynamic crossovers for supercritical AdS black holes }}

\vspace{10pt}
 Zhong-Ying Fan$^{1\dagger}$

\vspace{10pt}
$^{1\dagger}${ Department of Astrophysics, School of Physics and Material Science, \\
 Guangzhou University, No. 230 Wai Huan Xi Rd, Guangzhou 510006, P.R. China }\\

\vspace{40pt}

\underline{ABSTRACT}
\end{center}
Ising symmetry typically emerges in the critical domain between liquid-gas phases. Universality of this property imposes strong constraints on the behavior of thermodynamic crossovers for supercritical fluids. In this work, we develop a novel approach to investigate the crossover lines for supercritical AdS black holes using Lee-Yang phase transition theory. We analytically continue Lee-Yang zeros into the complex plane within the supercritical region by keeping a modular pressure real. Consequently, we obtain a pair of complex crossover lines, which exhibit universal scalings and manifest the emergent Ising symmetry in the complex phase space. The real crossover lines are defined by projecting the complex crossovers onto the real phase space. As a result, the phase diagram above the critical point is divided into three distinct regimes: liquid-like, indistinguishable and gas-like states, in sharp contrast to scenarios featuring only a single crossover line.

\vfill {\footnotesize  Email: fanzhy@gzhu.edu.cn\,.}

\thispagestyle{empty}

\pagebreak

\tableofcontents
\addtocontents{toc}{\protect\setcounter{tocdepth}{2}}




\section{Introduction}

Standard textbook knowledge posits that distinct liquid  and gas phases cease to exist in a supercritical fluid beyond the critical point. However, this single-phase scenario fails to account for the emergence of the condensed phases along a path traversing the critical point. Hence, it is generally proposed that a supercritical crossover line exists, dividing the supercritical region into liquid-like and gas-like states above the critical point. Several candidates for the crossover line have been proposed, including the Widom line \cite{xu2005relation, simeoni2010widom, ruppeiner2012thermodynamic, luo2014behavior, corradini2015widom, gallo2014widom, de2021widom}, the Frenkel line \cite{brazhkin2012two, yoon2018two, bolmatov2014structural, bolmatov2015frenkel, fomin2018dynamics, yang2015frenkel} and others \cite{fisher1969,PhysRevE.51.3146,tarazona2003,nishikawa1995,nishikawa1998,ploetz2019,woodcock2013}. Each of these definitions is grounded in distinct theoretical considerations. Among them, the Widom line is perhaps most widely accepted. However, its definition (and analogous generalizations \cite{Li_2024,Wang:2025ctk}) remains ambiguous. In practice, it is usually defined via the extremum of the isobaric heat capacity or the isothermal compressibility. While this is useful in many cases, there are no a priori rules which guarantee the existence of a global extremum of the isobaric heat capacity in general quantum fluids. 

This leads to a conceptual challenge in physics. For example, charged AdS black holes undergo a small-large black hole transition in the extended phase space, which is analogous to the liquid-gas transition of the Van der Waals (VdW) fluids \cite{Kubiznak:2012wp}. However, unlike in VdW fluids, the extremum of the isobaric heat capacity is strictly local, rendering the conventional definition of the Widom line inapplicable. To resolve this issue, the authors in \cite{Xu:2025jrk} employ Lee-Yang phase transition theory to analyze the phase diagram of the AdS black holes. In the supercritical region, by analytically continuing  Lee-Yang zeros (defined via singularities of the response functions) into the complex plane while keeping the pressure real, they derive the Widom line by projecting a complex crossover line onto the real phase space. Remarkably, this definition avoids the aforementioned ambiguity.

However, a significant issue arises in \cite{Xu:2025jrk}: all the complex zeros stay within the unit circle, resulting in a single crossover line in the supercritical region. On one hand, this partially contradicts the spirit of Lee-Yang theorem, which states that the roots of the grand partition function always lie on the unit circle \cite{Yang:1952be,Lee:1952ig}. Accordingly the Lee-Yang zeros located inside and outside the unit circle should be interpreted as the boundaries of distinct phases. For instance, in the subcritical region, the spinodal lines (corresponding to the zeros on the positive real axis) demarcate the boundaries of stable liquid-gas phases, respectively. One would naturally anticipate analogous behavior in the supercritical region. On the other hand, the existence of a single crossover line is essentially inconsistent with  the emergent Ising symmetry observed in critical phenomena \cite{Cui:2025ovm}. The latter implies that properly defined crossovers (whether subcritical or supercritical) should consist of a pair of lines, exhibiting a $Z_2$ symmetry with respect to the critical isochore. This serves as  a stringent constraint. Intriguingly, the spinodal lines do exhibit the emergent symmetry within the subcritical region. This inspires us to consider that a proper analytical continuation of the Lee-Yang zeros may yield thermodynamic crossovers in the supercritical region, which manifest the emergent symmetry as well. The answer will become clear once universal behaviors of Lee-Yang zeros in the critical regime is elucidated.


\section{Emergent Ising symmetry in criticality}

It turns out that for general quantum fluids, there exists a hidden symmetry between the condensed phases \cite{Cui:2025bfr}.  The algebraic conditions for the coexistence line implies that the functional relation between the specific volumes of the liquid-gas phases obeys
\be z_l=\varphi(z_g)\,,\qquad z_g=\varphi(z_l) \,,\label{zlg}\ee
where $z=v/v_c$ is the reduced specific volume and the subscripts $l\,,g$ stands for the liquid and the gas phases, respectively. The function $\varphi$ is called self-reciprocal because of $\varphi=\varphi^{-1}$. This nice property has remarkable consequences. Firstly, it enables us to solve the coexistence line analytically for a variety of cases, including the celebrated VdW model \cite{Cui:2025bfr}. Secondly, it implies that under suitable conditions, Ising symmetry typically emerges in the critical domain. Assume the function $\varphi$ is of $\mathcal{C}^1$ at least. Expanding the function $\varphi$ near the critical point yields to leading order 
\be \omega_g=-\omega_l \,,\label{orderz2}\ee
where $\omega=z-1$ is the order parameter. 
 It is known that the result holds for the Ising model because of the global symmetry of the Hamiltonian. Yet its universality and physical consequences are unfortunately unexplored  in literature.  

It turns out that the relation Eq. (\ref{orderz2}) constrains the equation of state to be 
\be  \tilde p=\sum_{0\leq i\leq [\ft{\delta-1}{2}]} b_{2i+1}\,\tau^{\beta(\delta-2i-1)}\,\omega^{2i+1}+\cdots \,,\label{modularp}\ee
where $p=P/P_c$ is the dimensionless reduced pressure, $\tilde p=p-p(\omega_c\,,\tau)$ defines a modular pressure obtained by subtracting the critical isochore $p(\omega_c\,,\tau)$ and $\tau\equiv 1-T/T_c$. Here $\beta\,,\delta$ are the ordinary exponents and the dots stands for higher order terms beyond the scaling regime. Notably, this extends the well-established properties of the Ising model to the realm of general quantum fluids. We may identify $\omega\rightarrow M\,,\tilde p\rightarrow H$, where $M$ is magnetization and $H$ the magnetic field. In the Ising model, the symmetry is exact, corresponding to the self-reciprocal function $\varphi(x)=-x$. Nevertheless, the symmetry induced constraints  will play a comparable role in general quantum fluids. The existence of a pair of thermodynamic crossover lines serves as a prime example of this analogy.

Derivation of Eq. (\ref{modularp}) relies on several conditions \cite{Cui:2025ovm}.  Firstly, the scaling hypothesis states that in the scaling regime
\be\label{masterp} \tilde p=\omega^\delta\psi(\mu) \,,\qquad \psi(\mu)=\sum_{\delta\geq i\geq 1} b_i \mu^{\delta-i}  \,,\ee
where $\mu=\tau^\beta\omega^{-1}$.  Secondly, the equation of states should be regular at $\tau=0$ for $\omega\neq 0$ and $\omega=0$ for $\tau>0$. This requires the scaling function $\psi(\mu)$ being truncated at the order $\mu^\delta$, which defines the relevant terms in the scaling regime.  Finally, the system under consideration should be thermodynamically stable so that $\fft{\partial^3p}{\partial\omega^3}\Big|_{\tau=0}<0$.
This implies that the exponent $\delta$ should be an odd integer and $\psi(0)<0$.

Combining these conditions with Eq. (\ref{orderz2}), it follows that only even powers of $\mu$ can survive in the scaling function $\psi$ and hence leads to Eq. (\ref{modularp}) inevitably. This clarifies a longstanding issue in textbooks: why even powers of $\omega$ can be dropped in the critical domain. As a byproduct, the fact that the scaling function $\psi$ is even will play a pivotal role in the critical scalings of Lee-Yang zeros.

\section{Universal scalings of complex crossover lines}

The Lee-Yang zeros are primarily defined as the zeros of the grand partition function $Z$ or equivalently, the singularities of the Gibbs free energy ($G=-T\,\mm{ln}Z$) \cite{Yang:1952be,Lee:1952ig}. This concept was extended in \cite{Xu:2025jrk} to the non-analyticities of $G$, corresponding to the singularities of response functions such as the isobaric heat capacity or isothermal compressibility. In this work, by Lee-Yang zeros, we refer to the same meanings of \cite{Xu:2025jrk}.

Consider the subcritical region at first. Using the scaling form Eq. (\ref{masterp}), we evaluate the dimensionless reduced isothermal compressibility
\be \kappa_T=-\fft{\partial \omega}{\partial p}\Big|_T=\fft{\omega^{1-\delta}}{\mu\,\psi'(\mu)-\delta\,\psi(\mu)} \,.\ee 
The Lee-Yang zeros are determined by
\be  \mu\,\psi'(\mu)-\delta\,\psi(\mu)=0\,.\label{masterzero}\ee
In the subcritical region, the zeros are real and positive definite, demarcating the boundaries of liquid-gas phases. Clearly, since $\psi$ contains only even powers of $\mu$, Eq. (\ref{masterzero}) admits the $Z_2$ symmetric solutions of the form
\be  \tilde p_\pm=\pm c_1\, \tau^{\beta\delta}\,,\qquad \omega_\pm=\pm c_2\, \tau^\beta \,,\label{solcrit}\ee 
where   the subscript ``$\pm$" stands for the liquid/gas phases and $c_1\,,c_2$ are two scale factors which are material dependent, determined by
\be c_2^{-1}\psi'(c_2^{-1})-\delta\,\psi(c_2^{-1})=0\,,\qquad c_1=c_2^\delta\psi(c_2^{-1}) \,.\ee
Here we choose $c_2>0$ whereas the sign of $c_1$ depends on the sign of the scaling function $\psi$, which is however path dependent. Generally we assume that for Lee-Yang zeros $\psi>0$ in the subcritical region and $\psi<0$ in the supercritical region, respectively. The change in sign of the scaling function across the critical point results from the fact that the transition occurs only in the subcritical region. 
This predicts that the spinodal lines exhibit universal scalings and manifest an emergent Ising symmetry in the critical domain. In this sense, they are better crossover lines than the coexistence line in the subcritical region. Previously, the same conclusion was drawn owing to metastable states of supercooling or superheating, existing in the regions between the spinodal lines and the coexistence line \cite{Xu:2025jrk}. Here we view it as a great physical impact of the symmetry constraint.

In the supercritical region, the scaling function will generally be different, say $\tilde p=\omega^\delta \tilde\psi(\mu)$ since now $\tau=T/T_c-1$. However, according to Eq. (\ref{modularp}), the function $\tilde{\psi}$ should still be even in the critical domain so we will omit the tilde for simplicity. The Lee-Yang zeros are still determined by Eq. (\ref{masterzero}) but now
no real solutions exist since no phase transition occurs in the supercritical region. However, an important lesson from the theory developed above is that the modular pressure $\tilde p$ plays a pivotal role in criticality. This motivates us to keep $\tilde p$ (rather than the pressure $p$ itself) real when analytically continuing the Lee-Yang zeros into the complex plane. 

Again, since the scaling function $\psi$ is even,  the $Z_2$ symmetric solution of the form Eq. (\ref{solcrit}) still  exists  but now the temperature and the scale factors $c_1\,,c_2$ become complex.
This defines  a novel complex scaling regime with real critical exponents. Notably, the zeros $\omega_+$ ($\omega_-$) are distributed inside (outside) the unit circle and hence could be interpreted as describing the crossover line of the liquid-like (gas-like) states, analogous to the spinodal lines case in the subcritical region.

Finally, we follow \cite{Xu:2025jrk} and define the physical crossover lines by projecting the complex crossovers Eq. (\ref{solcrit}) onto the real phase space within the supercritical region. In the critical domain, the complex Lee-Yang zeros approach to the critical point and hence $\mm{Re}(\tau)\propto |\tau|\,,\mm{Re}(\omega_\pm)\propto |\omega_\pm|$. Therefore the real crossover lines exhibit the universal scalings 
\be \tilde{p}\propto \pm \Big(\mm{Re}(\tau)\Big)^{\beta\delta}\,,\qquad \mm{Re}(\omega_\pm)\propto \mp \Big( \mm{Re}(\tau) \Big)^\beta \,.\ee
Clearly, the presence of two crossover lines will divide the phase diagram above the critical point into three distinct regimes: liquid-like, indistinguishable and gas-like states. This stands in sharp contrast to scenarios featuring only a single crossover line. It implies that within a supercritical fluid, liquid-like and gas-like states could be defined only in a limited sense according to the emergent symmetry. A similar situation occurs in the subcritical region. In that case, while the coexistence line strictly separates the phase space into liquid and gas phases, the regions between it and the spinodal lines correspond to metastable states of supercooling or superheating. In comparison, the precise meaning of the indistinguishable states in the supercritical region remains unclear in current work and certainly deserves further studies.

\section{Charged AdS black holes}

Consider holographic fluids dual to charged AdS black holes. 
The equation of states in the extended phase space reads \cite{Kubiznak:2012wp}
\be P=\fft{T}{v}-\fft{1}{2\pi v^2}+\fft{2Q^2}{\pi v^4} \,,\ee
where $P=-\Lambda/8\pi$ is the thermodynamic pressure and $v=2r_h$ stands for the specific volume of black hole molecules. Here $\Lambda$ is the cosmological constant, $r_h$ the horizon radius and $Q$ the electric charge of the black holes. In the extended phase space, the small-large black hole transition in many respects is analogous to the liquid-gas transition of the VdW fluids. The critical point occurs at 
\be
v_{c}=2\sqrt{6}\,Q\,,\quad T_c=\fft{\sqrt{6}}{18\pi Q}\,,\quad P_c=\fft{1}{96\pi Q^2}\,.
\ee
For convenience, we work with the reduced quantities $p=P/P_c\,,t=T/T_c\,,z=v/v_c$. We obtain the dimensionless Gibbs free energy $g=G/G_c=(3+6z^2-pz^4)/8z$ and the isothermal compressibility 
\be \kappa_T=-\fft{3(z-1)z^4}{(3\tilde{p}-5)z^4+12z^3-6z^2-4z+3}  \,.\ee 
It turns out that below the critical point, the coexistence line can be solved analytically as \cite{Spallucci:2013osa}
\be t=\sqrt{ \fft{p( 3-\sqrt{p} )}{2} }  \,.\ee 
This follows directly from the self-reciprocal property of the black holes and can be generalized to diverse dimensions \cite{Cui:2025bfr}.
\begin{figure}
\centering
\includegraphics[width=215pt]{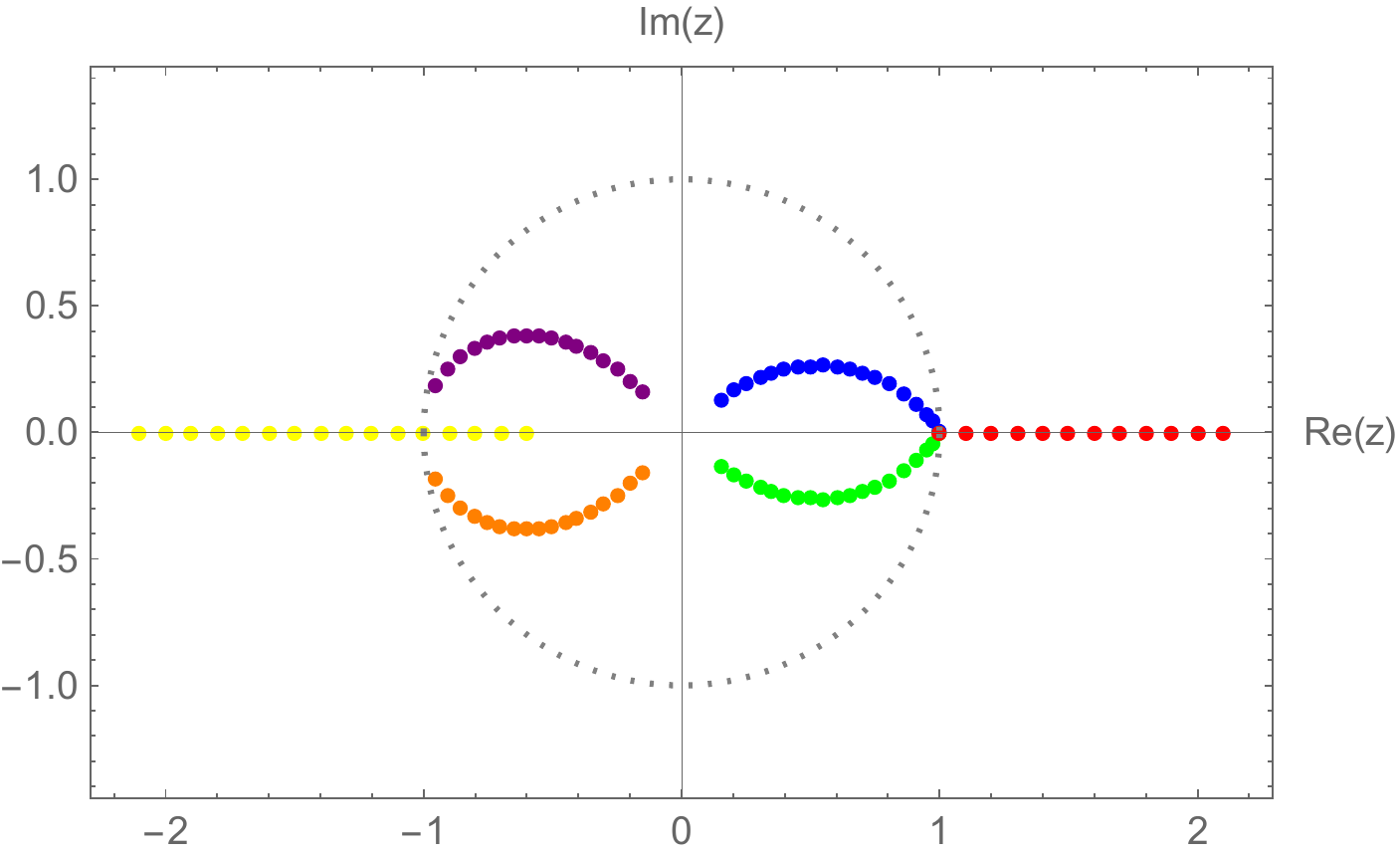}
\includegraphics[width=215pt]{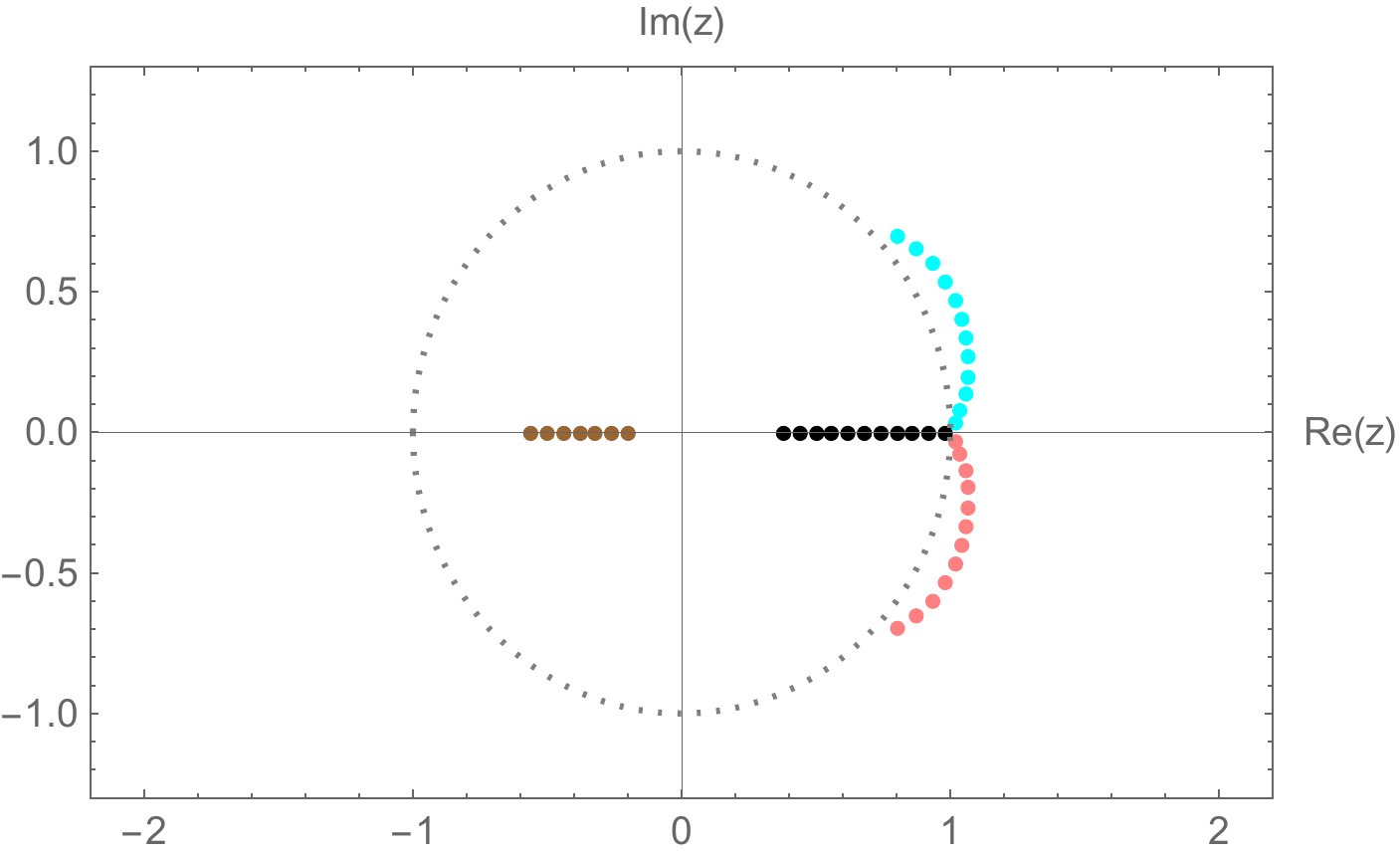}
\caption{The distribution of Lee-Yang zeros for $\tilde p>0$ (left) and $\tilde p<0$ (right). In both panels, the roots on the real axis correspond to $p<1$ whereas the complex roots correspond to $\mm{Re}(p)\geq 1$.   }
\label{roots}
\end{figure}

The Lee-Yang zeros are defined by singularities of the response function and hence are determined by
\be (3\tilde{p}-5)z^4+12z^3-6z^2-4z+3=0 \,.\label{adszero}\ee
Clearly, there exists four roots to Eq. (\ref{adszero}) for a given $\tilde p$. Depending on the sign of $\tilde p$, the distribution of the roots differs significantly, as depicted in the Fig. \ref{roots}. Since roots with a negative real part are unphysical, we restrict our discussion to those lying in the first and fourth quadrants, including those on the positive real axis. 

The positive real roots to Eq. (\ref{adszero}) describe the small (large) black hole states for $\tilde p>0$ ($\tilde p<0$) in the subcritical region. In fact, the distribution of these roots in our approach just presents an unconventional form for the spinodal lines $S^\pm$, which are usually expressed as
\be p=\fft{2z^2-1}{z^4}\,,\qquad t=\fft{3z^2-1}{2z^3} \,.\label{spinodal}\ee
In the scaling regime, these lines $S^\pm$ exhibit the critical scalings 
\be \tilde{p}_\pm= \pm\fft{16\sqrt{6}}{27}\,\tau^{3/2}\,,\qquad \omega_\pm=\pm\fft{\sqrt{6}}{3}\,\tau^{1/2} \,.\label{realscaling}\ee
Here it is worth emphasizing that the sign ``$+$'' (``$-$") stands for the small (large) black holes. This might be surprising but nothing is wrong here. The issue can be attributed to the behavior of the scaling function $\psi$ in the subcritical region. For the charged AdS black holes $\psi(\mu)=-\fft 43+\fft{8}{3}\mu^2$. While 
$\psi(0)<0$ according to the stability condition, the sign of $\psi$ is path dependent in the scaling regime. For example, along the spinodal lines $\mu=\sqrt{6}/2$ so that $\psi=8/3>0$. Considering the transition through an isobaric process, the line $S^+$ ($S^-$) may be interpreted as the boundary of super-cooled large (super-heated small) black holes.

In the supercritical region, there exists no real solutions to Eq. (\ref{adszero}). We analytically continue the Lee-Yang zeros into the complex plane by keeping the modular pressure $\tilde p$ real. This yields a complex temperature and a complex pressure because of $\mm{Im}(p)\propto \mm{Im}(t)$. The roots in the first and fourth quadrants in Fig. \ref{roots} are complex conjugates and hence are physically equivalent. We choose those in the first quadrant so that $\mm{Im}(t)>0$ for $\tilde p>0$ and $\mm{Im}(t)<0$ for $\tilde p<0$. This gives rise to a pair of complex crossover lines in the supercritical region and consequently leads to a complex phase diagram. 

\begin{figure}
\centering
\includegraphics[width=270pt]{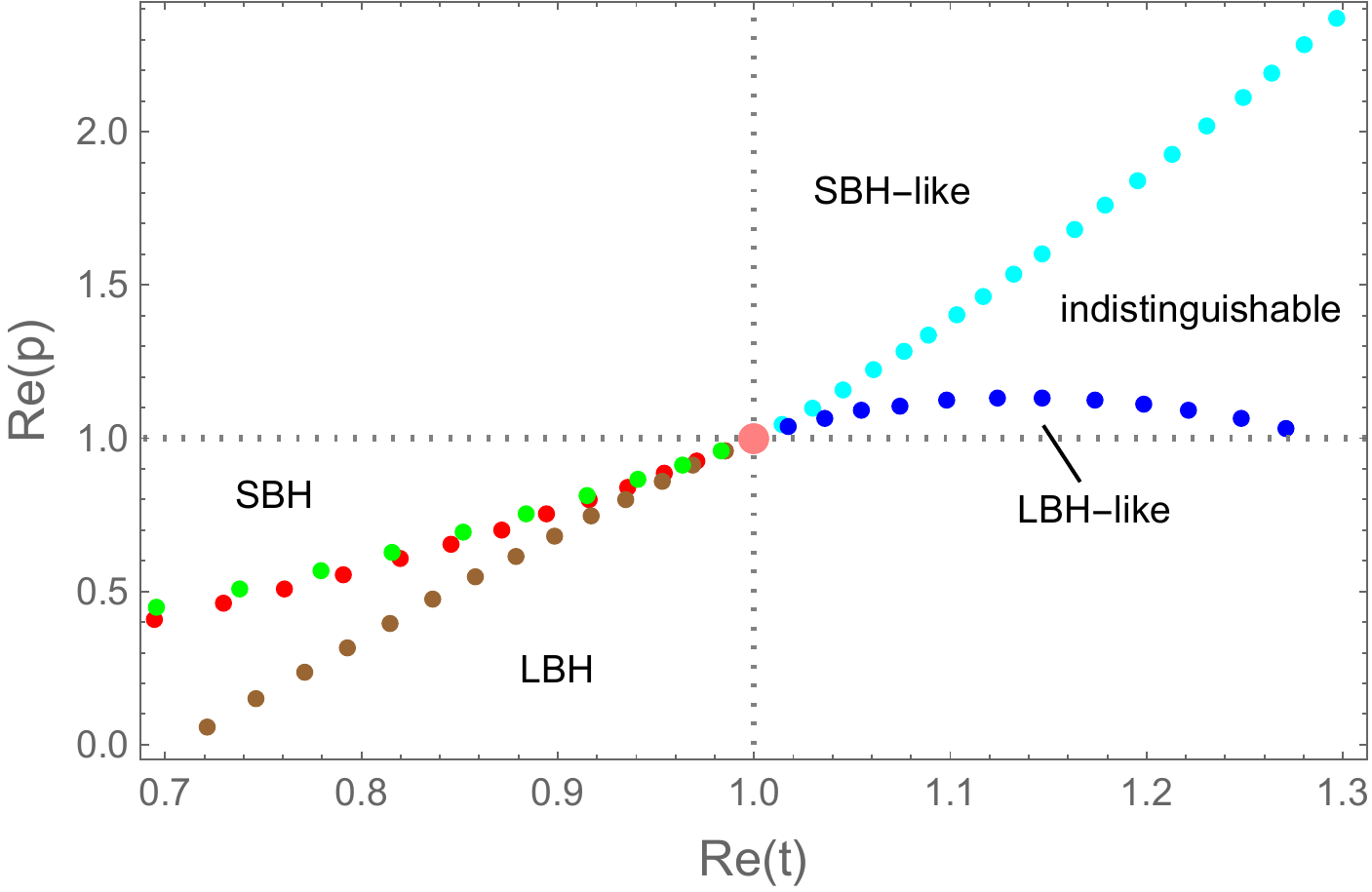}
\caption{The real phase diagram of charged AdS black holes. The dot plots are: green (spinodal line $S^+$), red (coexistence line), brown (spinodal line $S^-$), pink (critical point), cyan (crossover line $W^+$) and blue (crossover line $W^-$). }
\label{phase}
\end{figure}

Notably, since the roots for $\tilde p>0$ ($\tilde p<0$) stay inside (outside) the unit circle, they can be interpreted as marking the boundaries of the small black hole (SBH)-like and large black hole (LBH)-like states, respectively, an analogy to the subcritical case. The supercritical crossovers $W^\pm$ in the real phase space are obtained by projecting these complex crossover lines onto the real phase space within the supercritical region, as depicted in Fig. \ref{phase}. Clearly, the supercritical phase space is divided into three distinct regimes: SBH-like, indistinguishable and LBH-like states. Notice that the crossover lines $W^\pm$ terminate at specific points (e.g.,  $W^-$ line terminates at $\mm{Re}(p)=1$). Similar behavior appears  widely for various candidates of thermodynamic crossover lines \cite{Li_2024,Wang:2025ctk,Cui:2025ovm}, including the ordinarily defined Widom line. However, resolving this issue is not the aim of the present work since our primary interest lies in the critical domain, where the emergent symmetry imposes strong constraints on the behavior of the crossover lines.  

In the complex scaling regime, the Lee-Yang zeros exhibit the scaling behavior
\be \tilde{p}_\pm= \pm\fft{16\sqrt{6}i}{27} \tau^{3/2}\,,\qquad \omega_\pm=\mp \fft{\sqrt{6}i}{3}\tau^{1/2} \,.\ee
 Notice that reality of $\tilde p$ constrains the azimuthal angle $\theta$ of the temperature to be $\theta=2k\pi+\pi/3$, where $k$ is odd. It is straightforward to see that the real crossover lines $W^\pm$ exhibit the critical scalings
\be \tilde{p}_\pm= \pm\fft{64\sqrt{6}}{27} \Big(\mm{Re}(\tau)\Big)^{3/2}\,,\qquad \mm{Re}(\omega_\pm)=\mp\fft{\sqrt{3}}{3} \Big(\mm{Re}(\tau)\Big)^{1/2} \,.\ee
Here the sign of $\omega^\pm$ is attributed to the behavior of the scaling function in the supercritical region. One finds $\psi(\mu)=-\fft43-\fft{8}{3}\mu^2$, which is always negative definite. Physically, it simply reflects the fact that no first order transition occurs within the supercritical region.
\begin{figure}
\centering
\includegraphics[width=270pt]{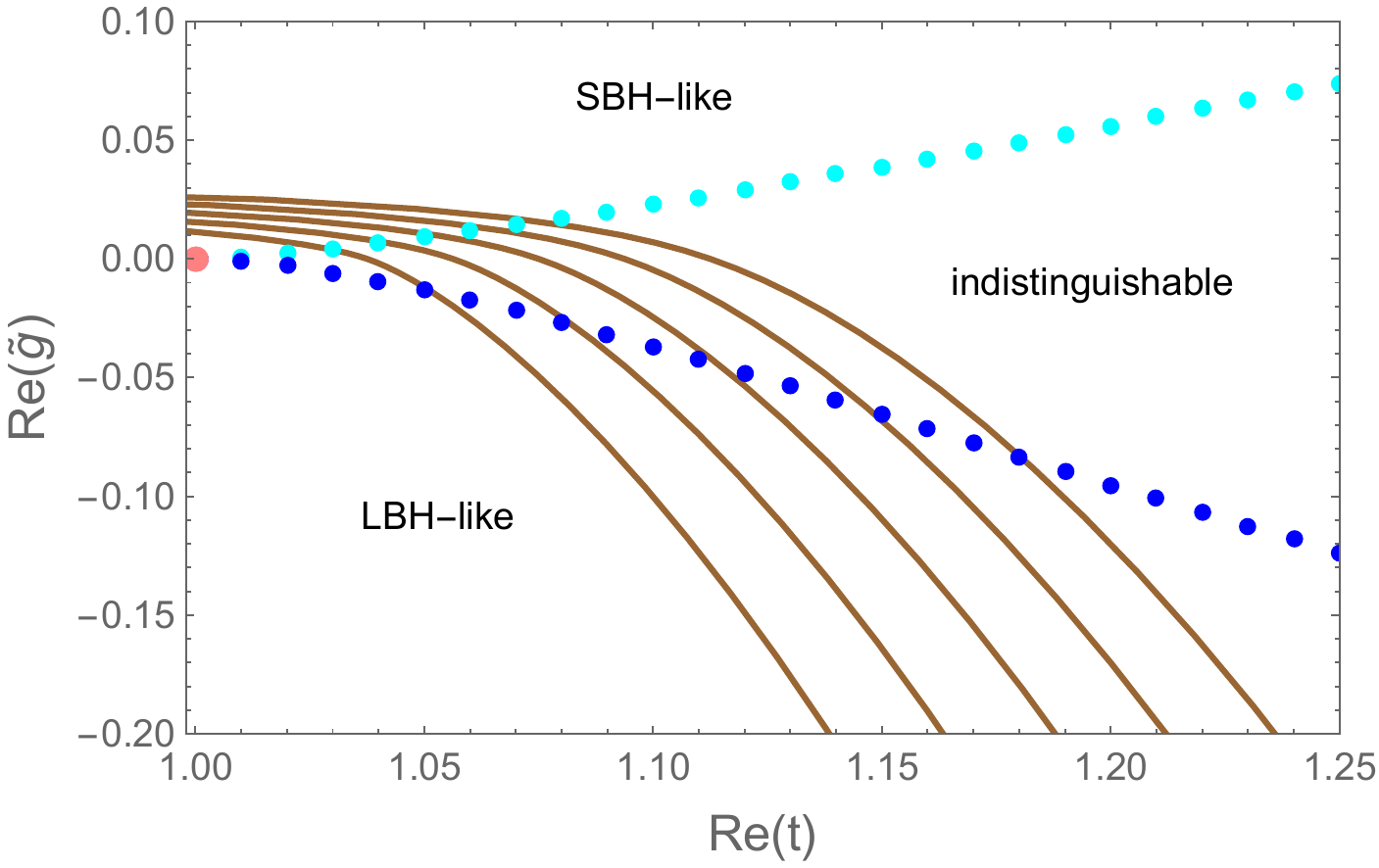}
\caption{The real phase diagram of supercritical AdS black holes on the $\tilde{g}-t$ plane. The dot plots take the same meanings as before. The modular Gibbs free energy in the real phase space is represented by solid lines and $p=1.1\,,1.15\,,1.2\,,1.25\,,1.3$, from bottom to top. }
\label{phasegt}
\end{figure}

The various supercritical phases can also be distinguished by the behavior of the Gibbs free energy $g$ upon projecting the crossover lines $W^\pm$ onto the real $g-t$ plane, as shown in Fig. \ref{phasegt}. Here we introduce a modular free energy as $\tilde g\equiv g-g(z_c\,,t)$. In general, statistical relations imply $\tilde g\propto \tilde p$. Consequently, the modular free energy exhibits the universal scalings $\tilde{g}\propto \pm \big(\mm{Re}(\tau)\big)^{\beta\delta}$ in the supercritical region along the crossover lines $W^\pm$. Upon crossing the critical point, the same scalings of $\tilde g$ reemerges on the spinodal lines $S^\pm$, which demarcate the boundaries of the stable liquid-gas phases. This establishes a symmetry-based framework for the emergence of condensed phases below the critical point. The situation is reminiscent of the Ising model, despite that the Ising symmetry here is approximate and no structural order changes below the critical point.


\section{Conclusions}
Many studies propose the existence of a supercritical crossover line, which divides the phase diagram above the critical point into two regimes with disparate physical properties. 
However, most of these works fail to account for the physical constraints imposed by the emergent Ising symmetry in the critical domain, generally resulting in a single crossover line. In contrast, we have established that emergence of the Ising symmetry is a universal feature of critical phenomena, rather than a peculiarity of the Ising model itself. This universality suggests that properly defined crossover lines should consist of a pair of lines, which are $Z_2$ symmetric (with respect to the critical isochore) within the critical domain. 

In this work, we developed a novel approach to study thermodynamic crossover lines for supercritical AdS black holes employing Lee-Yang phase transition theory. We have shown how to properly analytically continue Lee-Yang zeros into the complex plane within the supercritical region. Consequently, 
we obtain a pair of complex crossover lines, which exhibit universal scalings and manifest the emergent Ising symmetry in the complex phase space. The real crossover lines are then obtained by projecting the complex crossovers onto the real phase space.  As a result, the supercritical phase space is divided into three  distinct regimes: liquid-like, indistinguishable and gas-like states, rather than the two regimes characteristic of scenarios involving only a single crossover line. A similar situation occurs in the subcritical region, where the spinodal lines separate the phase space into liquid, metastable and gas phases, respectively. Both cases stem from the influence of emergent symmetry. To our knowledge, our work presents the first unambiguous definition of thermodynamic crossovers, consistent with the emergent symmetry in critical phenomena.


\section*{Acknowledgments}

The author thanks Hong-Ming Cui for valuable discussions in early stage of this work. Z.Y. Fan was supported in part by the National Natural Science Foundations of China with Grant No. 11873025.

\appendix

\newpage

\bibliographystyle{utphys}
\bibliography{reference}

@article{xu2005relation,
  title={Relation between the Widom line and the dynamic crossover in systems with a liquid--liquid phase transition},
  author={Xu, Limei and Kumar, Pradeep and Buldyrev, Sergey V and Chen, S-H and Poole, Peter H and Sciortino, Francesco and Stanley, H Eugene},
  journal={Proceedings of the National Academy of Sciences},
  volume={102},
  number={46},
  pages={16558--16562},
  year={2005},
  publisher={National Acad Sciences}
}

@article{simeoni2010widom,
  title={The Widom line as the crossover between liquid-like and gas-like behaviour in supercritical fluids},
  author={Simeoni, GG and Bryk, T and Gorelli, FA and Krisch, M and Ruocco, Giancarlo and Santoro, M and Scopigno, Tullio},
  journal={Nature Physics},
  volume={6},
  number={7},
  pages={503--507},
  year={2010},
  publisher={Nature Publishing Group}
}

@article{ruppeiner2012thermodynamic,
  title={Thermodynamic geometry, phase transitions, and the Widom line},
  author={Ruppeiner, George and Sahay, Anurag and Sarkar, Tapobrata and Sengupta, Gautam},
  journal={Physical Review E},
  volume={86},
  number={5},
  pages={052103},
  year={2012},
  publisher={APS}
}

@article{luo2014behavior,
  title={Behavior of the Widom line in critical phenomena},
  author={Luo, Jiayuan and Xu, Limei and Lascaris, Erik and Stanley, H Eugene and Buldyrev, Sergey V},
  journal={Physical Review Letters},
  volume={112},
  number={13},
  pages={135701},
  year={2014},
  publisher={APS}
}

@article{corradini2015widom,
  title={The Widom line and dynamical crossover in supercritical water: Popular water models versus experiments},
  author={Corradini, D and Rovere, M and Gallo, P},
  journal={The Journal of Chemical Physics},
  volume={143},
  number={11},
  pages={114502},
  year={2015},
  publisher={AIP Publishing LLC}
}

@article{gallo2014widom,
  title={Widom line and dynamical crossovers as routes to understand supercritical water},
  author={Gallo, P and Corradini, D and Rovere, M},
  journal={Nature communications},
  volume={5},
  number={1},
  pages={1--6},
  year={2014},
  publisher={Nature Publishing Group}
}

@article{de2021widom,
  title={Widom line of real substances},
  author={de Jes{\'u}s, EN and Torres-Arenas, J and Benavides, AL},
  journal={Journal of Molecular Liquids},
  volume={322},
  pages={114529},
  year={2021},
  publisher={Elsevier}
}

@article{Li_2024,
   title={Thermodynamic crossovers in supercritical fluids},
   author={Li, Xinyang and Jin, Yuliang},
   journal={Proceedings of the National Academy of Sciences},
   volume={121},
   number={18},
   pages={1091-6490},
   year={2024},
   month=apr,
   publisher={Proceedings of the National Academy of Sciences}
}

@article{Wang:2025ctk,
    author = "Wang, Shoucheng and Li, Xinyang and Jin, Yuliang and Li, Li",
    title = "{Analogous supercritical crossovers in black holes and water}",
    eprint = "2506.10808",
    archivePrefix = "arXiv",
    primaryClass = "gr-qc",
    month = "6",
    year = "2025"
}

@article{Kubiznak:2012wp,
    author = "Kubiznak, David and Mann, Robert B.",
    title = "{P-V criticality of charged AdS black holes}",
    eprint = "1205.0559",
    archivePrefix = "arXiv",
    primaryClass = "hep-th",
    doi = "10.1007/JHEP07(2012)033",
    journal = "JHEP",
    volume = "07",
    pages = "033",
    year = "2012"
}

@article{Xu:2025jrk,
    author = "Xu, Zhen-Ming and Mann, Robert B.",
    title = "{Thermodynamic Supercriticality and Complex Phase Diagram for the AdS Black Hole}",
    eprint = "2504.05708",
    archivePrefix = "arXiv",
    primaryClass = "gr-qc",
    doi = "10.1103/c39y-zcz6",
    journal = "Phys. Rev. Lett.",
    volume = "136",
    number = "4",
    pages = "041402",
    year = "2026"
}

@article{Yang:1952be,
    author = "Yang, Chen-Ning and Lee, T. D.",
    title = "{Statistical theory of equations of state and phase transitions. 1. Theory of condensation}",
    doi = "10.1103/PhysRev.87.404",
    journal = "Phys. Rev.",
    volume = "87",
    pages = "404--409",
    year = "1952"
}

@article{Lee:1952ig,
    author = "Lee, T. D. and Yang, Chen-Ning",
    title = "{Statistical theory of equations of state and phase transitions. 2. Lattice gas and Ising model}",
    doi = "10.1103/PhysRev.87.410",
    journal = "Phys. Rev.",
    volume = "87",
    pages = "410--419",
    year = "1952"
}

@article{Cui:2025ovm,
    author = "Cui, Hong-Ming and Fan, Zhong-Ying",
    title = "{Emergent Ising symmetry and supercritical fluids}",
    eprint = "2509.17038",
    archivePrefix = "arXiv",
    primaryClass = "cond-mat.stat-mech",
    doi = "10.1007/JHEP02(2026)145",
    journal = "JHEP",
    volume = "02",
    pages = "145",
    year = "2026"
}

@article{Cui:2025bfr,
    author = "Cui, Hong-Ming and Fan, Zhong-Ying",
    title = "{Analytical approach to criticality of AdS black holes}",
    eprint = "2506.20959",
    archivePrefix = "arXiv",
    primaryClass = "gr-qc",
    doi = "10.1007/JHEP09(2025)130",
    journal = "JHEP",
    volume = "09",
    pages = "130",
    year = "2025"
}

@article{Spallucci:2013osa,
    author = "Spallucci, Euro and Smailagic, Anais",
    title = "{Maxwell's equal area law for charged Anti-deSitter black holes}",
    eprint = "1305.3379",
    archivePrefix = "arXiv",
    primaryClass = "hep-th",
    doi = "10.1016/j.physletb.2013.05.038",
    journal = "Phys. Lett. B",
    volume = "723",
    pages = "436--441",
    year = "2013"
}

@article{brazhkin2012two,
  title={Two liquid states of matter: A new dynamic line on a phase diagram},
  author={Brazhkin, VV and Fomin, Yu D and Lyapin, AG and Ryzhov, VN and Trachenko, K},
  journal={Physical Review E},
  volume={85},
  number={3},
  pages={031203},
  year={2012},
  publisher={APS}
}

@article{yoon2018two,
title = {“Two-Phase” thermodynamics of the Frenkel Line},
author = {Yoon, Tae Jun and Ha, Min Young and Lee, Won Bo and Lee, Youn-Woo},
journal = {The Journal of Physical Chemistry Letters},
volume = {9},
number = {16},
pages = {4550-4554},
year = {2018},
publisher={ACS Publications}
}

@article{bolmatov2014structural,
  title={Structural evolution of supercritical CO2 across the Frenkel line},
  author={Bolmatov, Dima and Zav’Yalov, D and Gao, M and Zhernenkov, Mikhail},
  journal={The journal of physical chemistry letters},
  volume={5},
  number={16},
  pages={2785--2790},
  year={2014},
  publisher={ACS Publications}
}

@article{bolmatov2015frenkel,
  title={The Frenkel Line: a direct experimental evidence for the new thermodynamic boundary},
  author={Bolmatov, Dima and Zhernenkov, Mikhail and Zav’yalov, Dmitry and Tkachev, Sergey N and Cunsolo, Alessandro and Cai, Yong Q},
  journal={Scientific reports},
  volume={5},
  number={1},
  pages={1--10},
  year={2015},
  publisher={Nature Publishing Group}
}

@article{fomin2018dynamics,
  title={Dynamics, thermodynamics and structure of liquids and supercritical fluids: crossover at the Frenkel line},
  author={Fomin, Yu D and Ryzhov, VN and Tsiok, EN and Proctor, JE and Prescher, C and Prakapenka, VB and Trachenko, K and Brazhkin, VV},
  journal={Journal of Physics: Condensed Matter},
  volume={30},
  number={13},
  pages={134003},
  year={2018},
  publisher={IOP Publishing}
}

@article{yang2015frenkel,
  title={Frenkel line and solubility maximum in supercritical fluids},
  author={Yang, C and Brazhkin, VV and Dove, MT and Trachenko, K},
  journal={Physical Review E},
  volume={91},
  number={1},
  pages={012112},
  year={2015},
  publisher={APS}
}

@article{fisher1969,
title = {Decay of correlations in linear systems},
author = {Fisher, M.E. and Wiodm, B},
journal = {The Journal of Chemical Physics},
volume = {50},
pages = {3756},
year = {1969},
publisher={ACS Publications}
}

@article{PhysRevE.51.3146,
  title = {Location of the Fisher-Widom line for systems interacting through short-ranged potentials},
  author = {Vega, C. and Rull, L. F. and Lago, S.},
  journal = {Phys. Rev. E},
  volume = {51},
  issue = {4},
  pages = {3146--3155},
  numpages = {0},
  year = {1995},
  month = {Apr},
  publisher = {American Physical Society},
}

@article{tarazona2003,
  title = {The Fisher-Widom line for systems with low melting temperature},
  author = {Tarazona, P. and Chacon, E. and Velasco, E. },
  journal = {Molecular Physics},
  volume = {101},
  pages = {1595},
  year = {2003},
  publisher = {American Physical Society},
}

@article{nishikawa1995,
  title = {Correlation lengths and density fluctuations in supercritical states of carbon dioxide},
  author = {Nishikawa, K. and Tanaka, I.},
  journal = {Chemical physics letters},
  volume = {244},
  pages = {149},
  year = {1995},
 publisher = {American Physical Society},
}

@article{nishikawa1998,
  title = {Fluid behavior at supercritical states studied by small-angle x-ray scattering},
  author = {Nishikawa, K. and Morita, T.},
  journal = {Journal of supercritical fluids},
  volume = {13},
  pages = {143},
  year = {1998},
 publisher = {American Physical Society},
}

@article{ploetz2019,
  title = {Gas or liquid? the supercritical behavior of pure fluids},
  author = {Ploetz, E. A. and Smith, P. E.},
  journal = {The Journal of Physical Chemistry B},
  volume = {123},
  pages = {6554},
  year = {2019},
 publisher = {American Physical Society},
}

@article{woodcock2013,
  title = {Observations of a thermodynamic liquid-gas critical coexistence line and supercritical fluid phase bounds from percolation transition loci},
  author = {Woodcock, L. V.},
  journal = {Fluid Phase Equilibria},
  volume = {351},
  pages = {25},
  year = {2013},
 publisher = {American Physical Society},
}

\end{document}